\documentclass[a4paper,10pt,twoside]{cpc-hepnp}

\usepackage{multicol}
\usepackage{graphicx}
\usepackage{booktabs}
\usepackage{amssymb,bm,mathrsfs,bbm,amscd}
\usepackage[tbtags]{amsmath}
\usepackage{lastpage}

\newcommand{\seq}{\begin{subequations}}
\newcommand{\sen}{\end{subequations}}
\newcommand{\eq}{\begin{eqnarray}}
\newcommand{\en}{\end{eqnarray}}
\newcommand{\ra}{\rangle}

\def\dppm{D^\pm}
\def\dpp{D^+}
\def\dpm{D^-}
\def\dpmps{D^{\ast \mp}}
\def\dpps{D^{\ast +}}
\def\dpms{D^{\ast -}}

\def\d{D^0}
\def\db{\bar{D}^{0}}
\def\ds{D^{\ast  0}}

\def\dbs{\bar{D}^{\ast 0}}

\def\L2{\Lambda^2}

\def\jp{J/\psi}
\def\jpsi{J_\psi}

\begin{document}

\fancyhead[co]{\footnotesize T. Gutsche et al: Hadron Molecules}

\footnotetext[0]{Received 24 December 2009}

\title{Hadron Molecules\thanks{Supported by Deutsche Forschungsgemeinschaft Contract
No. FA67/31-2 and No. GRK683. This research is also part of the
European Community-Research Infrastructure Integrating Activity
"Study of Strongly Interacting Matter" (acronym HadronPhysics2,
Grant Agreement No. 227431), Russian President grant
``Scientific Schools''  No. 3400.2010.2, Russian Science and
Innovations Federal Agency contract No. 02.740.11.0238.}}

\author{%
      Thomas Gutsche $^{1)}$\email{gutsche@uni-tuebingen.de}%
\quad Tanja Branz
\quad Amand Faessler
\quad Ian Woo Lee
\quad Valery E. Lyubovitskij
}
\maketitle

\address{%
Institut f\"ur Theoretische Physik,
Universit\"at T\"ubingen,
Kepler Center for Astro and Particle Physics,\\
Auf der Morgenstelle 14, D--72076 T\"ubingen, Germany\\}

\begin{abstract}
We discuss a possible interpretation of the open charm mesons $D_{s0}^*(2317)$,
$D_{s1}(2460)$ 
and the hidden charm mesons $X(3872)$, $Y(3940)$ and $Y(4140)$ 
as hadron molecules.
Using a phenomenological Lagrangian approach we review the strong and radiative
decays of the $D_{s0}^* (2317)$ and $D_{s1}(2460)$ states.
The $X(3872)$ is assumed
to consist dominantly of molecular hadronic components with an additional small
admixture of a charmonium configuration. Determing the radiative
($\gamma J/\psi$ and $\gamma \psi(2s)$) and strong ($J/\psi 2\pi $ and $ J/\psi 3\pi$)
decay modes we show that present experimental observation is consistent with
the molecular structure assumption of the $X(3872)$. 
Finally we give evidence for
molecular interpretations of the $Y(3940)$ and $Y(4140)$ 
related to the observed strong decay modes $J/\psi + \omega$ or 
$J/\psi + \phi$ , respectively.
\end{abstract}



\begin{multicols}{2}

\section{Introduction}

The complexity of hadronic mass spectra allows for the possibility that existing and newly
observed hadrons can possibly be interpreted as hadronic molecules. Such an interpretation
is possible when the mass of the observed meson lies close to or slightly below the
threshold of a corresponding hadronic pair. In the following we focus on some
of the candidates in the heavy meson sector which are discussed and considered as such
hadronic bound states. 

The $X(3872)$ is one of the peculiar new meson resonances which was discovered
during the last years~\cite{Amsler:2008zz} and where its properties cannot
be simply explained in the context of conventional constituent quark models.
In the molecular approaches, the earliest given in Refs.~\citep{Voloshin:1976ap, Tornqvist:1993ng},
it is argued that the
$X(3872)$ can be identified with a weakly--bound hadronic molecule
whose constituents are $D$ and $D^\ast$ mesons. This natural interpretation
is due to the fact that its mass is very close to the
$\d \dbs$ threshold and hence is in analogy to the deuteron ---
a weakly--bound state of proton and neutron.

In the open charm sector the scalar
$D_{s0}^\ast(2317)$
and axial $D_{s1}(2460)$ mesons could be candidates for a scalar $DK$
and a axial $D^\ast K$ molecule because of a relatively small binding
energy of $\sim 50$ MeV. These states were discovered and confirmed just a few years ago by
the Collaborations BABAR~\cite{Aubert:2003fg}, CLEO~\cite{Besson:2003cp}
and Belle~\cite{Abe:2003jk}.
The simplest
interpretation of these states is that they are the missing $j_s = 1/2$
(the angular momentum of the $s$-quark) members of the $c \bar s$ $L=1$
multiplet. However, this standard quark model scenario is in
disagreement with experimental observation since the $D_{s0}^\ast(2317)$
and $D_{s1}(2460)$ states are narrower and their masses are lower when
compared to theoretical expectations
(see e.g. discussion in Ref.~\citep{Rosner:2006vc}).

The CDF Collaboration recently reported evidence
for a narrow near-threshold structure, termed the $Y(4140)$ meson,
in the $J/\psi \phi$ mass spectrum in exclusive $B^+ \to J/\psi \phi K^+$
decays with the mass $m_{Y(4140)} = 4143.0 \pm 2.9 ({\rm stat}) \pm
1.2 ({\rm syst})$ MeV
and natural width $\Gamma_{Y(4140)} = 11.7^{+8.3}_{-5.0} ({\rm stat})
\pm 3.7 ({\rm syst})$ MeV~\cite{Aaltonen:2009tz}.
As already stressed in \cite{Aaltonen:2009tz}, the new structure $Y(4140)$,
which decays to $J/\psi \phi$ just above the $J/\psi \phi$ threshold,
is similar to the previously discovered
$Y(3940)$~\cite{Abe:2004zs,Aubert:2007vj},
which decays to $J/\psi\omega$ near this respective threshold.
Both observed states, $Y(4140)$ and $Y(3940)$, are well
above the threshold for open charm decays. A conventional
$c\bar c$ charmonium interpretation is disfavored, since open charm decay
modes would dominate, while the $J/\psi \phi$ or $J/\psi \omega$ decay rates
are essentially negligible~\cite{Aaltonen:2009tz,Eichten:2007qx}.
This could be a signal for nonconventional structure of these $Y$
states.

In the following we give a compact overview of our recent calculations
related to the strong and electromagnetic decay properties of
the $D_{s0}^\ast(2317)$, $D_{s1}(2460)$, $X(3872)$, $Y(3940)$
and $Y(4140)$ mesons interpreted
as hadron molecules. Details can be found in the related publications
\cite{Faessler:2008vc,Faessler:2007us,Faessler:2007gv,Faessler:2007cu,
Dong:2009uf,Dong:2009yp,Lee:2009hy,Dong:2008gb,Branz:2009yt,Branz:2008cb}
on this topic.

\section{The method}

We first briefly discuss the formalism behind the study of hadronic
molecules. As an example we consider the $D_{s0}^{\ast\pm}(2317)$ meson
as a bound state of a $D$ and a $K$ meson.
We use the current results for the quantum numbers
of isospin, spin and parity: $I(J^P) = 0(0^+)$ and mass
$m_{D_{s0}^{\ast}} = 2.3173$~GeV~\cite{Amsler:2008zz}.
Our framework is based on an effective interaction
Lagrangian describing the coupling between the $D_{s0}^\ast(2317)$
meson and their constituents - $D$ and $K$ mesons:
\eq\label{Lagr_Ds0}
{\cal L}_{D_{s0}^\ast}(x) \, &=& \, g_{_{D_{s0}^\ast}} \,
D_{s0}^{\ast \, -}(x) \, \int\! dy \,
\Phi_{D_{s0}^\ast}(y^2) \cdot \nonumber \\ && D^T(x+w_K y) \,
K(x-w_D y) \, + \, {\rm H.c.}
\en
where $D$ and $K$ are the meson doublets and
$w_{ij} = \frac{m_i}{m_i + m_j}$ is a kinematic variable.
The correlation function $\Phi_{D_{s0}^\ast}$ characterizes the finite
size of the $D_{s0}^\ast(2317)$ meson as a $(DK)$ bound state and
depends on the relative Jacobi coordinate $y$ with
$x$ being the center of mass (CM) coordinate.
The Fourier transform of the correlation function is
$\tilde\Phi_{D_{s0}^\ast}(p_E^2)
\doteq \exp( - p_E^2/\Lambda^2_{D_{s0}^\ast})$, where $p_{E}$ is the
Euclidean Jacobi momentum.
Here $\Lambda_{D_{s0}^\ast}$
is a size parameter, which parametrizes the distribution of
$D$ and $K$ mesons inside the $D_{s0}^\ast$ molecule.
The $D_{s0}^\ast DK$ coupling constant $g_{D_{s0}^\ast}$
is determined by the compositeness
condition~\cite{Weinberg:1962hj,Efimov:1993ei},
which implies that the renormalization constant of the hadron
wave function is set equal to zero:
$Z_{D_{s0}^\ast} = 1 -
\Sigma^\prime_{D_{s0}^\ast}(m_{D_{s0}^\ast}^2) = 0$,
where
$\Sigma^\prime_{D_{s0}^\ast}$
is the derivative of the $D_{s0}^\ast$ meson mass operator.
The effective Lagrangian is the basis for the study of the
decay properties of the hadronic molecules. It defines the coupling of the
molecule to its constituents but also supplies the hadron loop connected
to the final state particles considered. The interaction of the constituents
with external particles is further described by effective Lagrangians
either adjusted by data or by theory (as for example by heavy hadron chiral
perturbation theory~\cite{Wise:1992hn,Colangelo:2003sa} (HHCHPT)). Further details of this procedure can be found
in Refs. \citep{Faessler:2008vc,Branz:2008cb}.

\section{Strong and radiative decays of $D_{s0}^*(2317)$ and
$D_{s1}(2460)$}

For the $D_{s0}^\ast(2317)$ we consider a possible interpretation
as a hadronic molecule - a bound state
of $D$ and $K$ mesons. Using an effective Lagrangian approach
we calculated the strong $D_{s0}^{\ast} \to D_s \pi^0$ and
radiative $D_{s0}^{\ast} \to D_s^{\ast} \gamma$
decays~\cite{Faessler:2007gv}. A new impact related to the $DK$ molecular structure
of the $D_{s0}^\ast(2317)$ meson is that the presence of $u(d)$
quarks in the $D$ and $K$ mesons gives rise to a direct strong
isospin-violating transition $D_{s0}^{\ast} \to D_s \pi^0$ (see Fig.1) in
addition to the decay mechanism induced by $\eta-\pi^0$ mixing
considered previously (see Fig.2).
\begin{center}
\includegraphics[width=6cm]{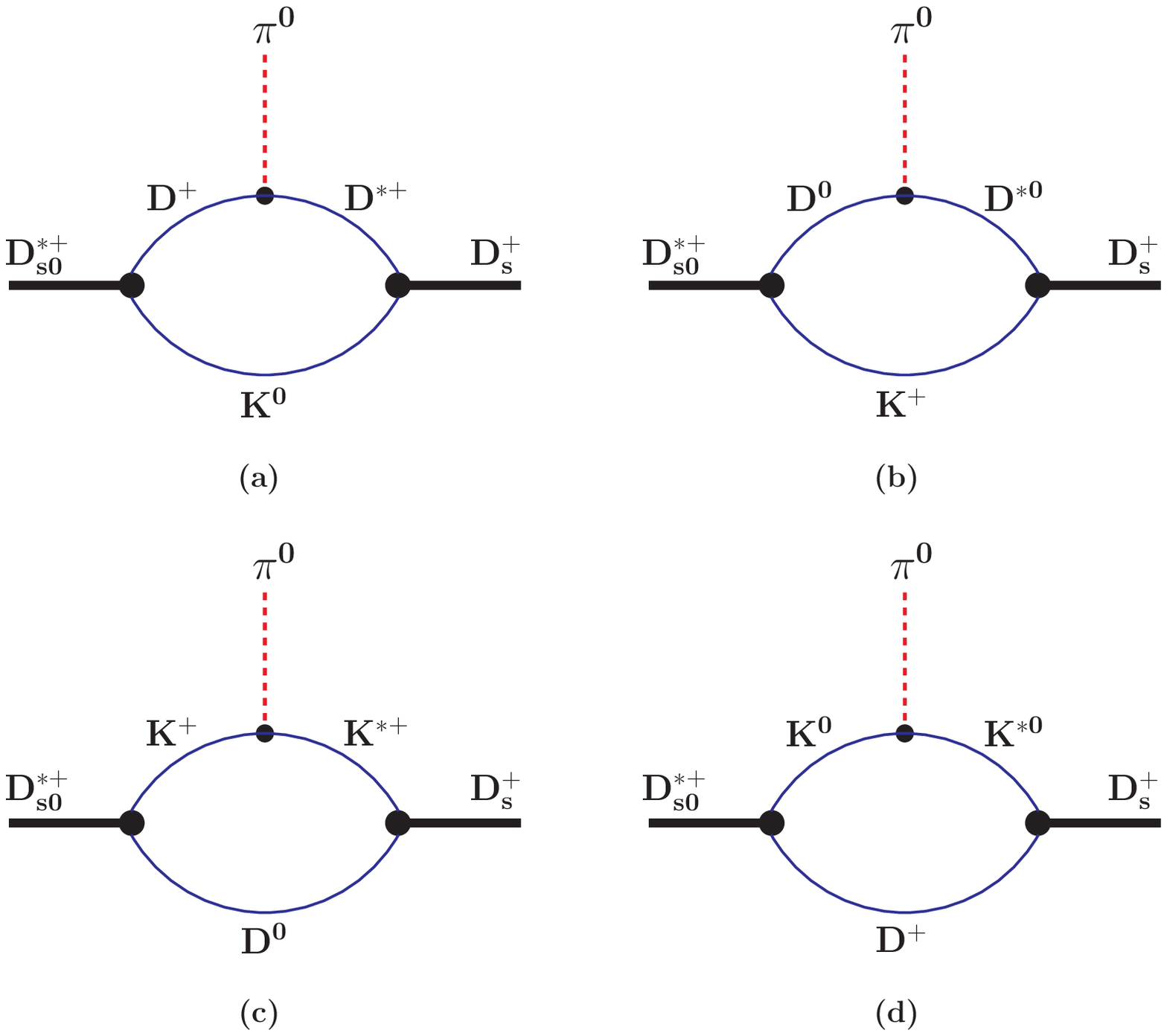}
\figcaption{\label{fig1} Direct isospin-violating transition for
$D_{s0}^{\ast} \to D_s \pi^0$}
\end{center}
We show~\cite{Faessler:2007gv} that the direct transition
dominates over the $\eta-\pi^0$ mixing transition in the
$D_{s0}^{\ast} \to D_s \pi^0$ decay.
\begin{center}
\includegraphics[width=6cm]{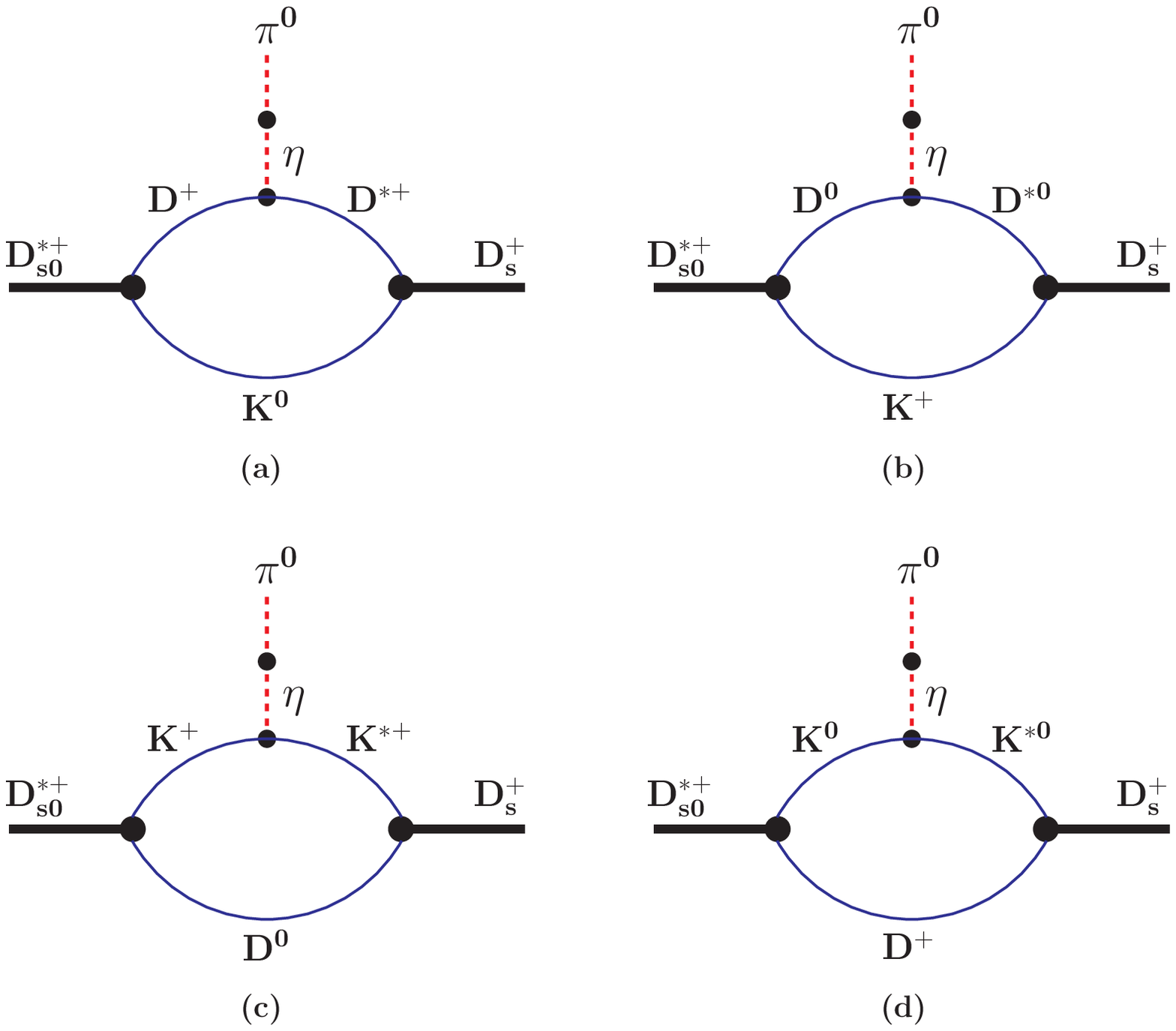}
\figcaption{\label{fig3} $\eta-\pi^0$ mixing transition for $D_{s0}^{\ast} \to D_s \pi^0$}
\end{center}
The radiative transition $D_{s0}^{\ast +} \to D_{s}^{\ast +} \gamma$
is generated by the loop diagrams of Fig.3, where the last graphs
have to be included to guarantee full gauge invariance for the case
of a non-local vertex. 
\begin{center}
\includegraphics[width=6.5cm]{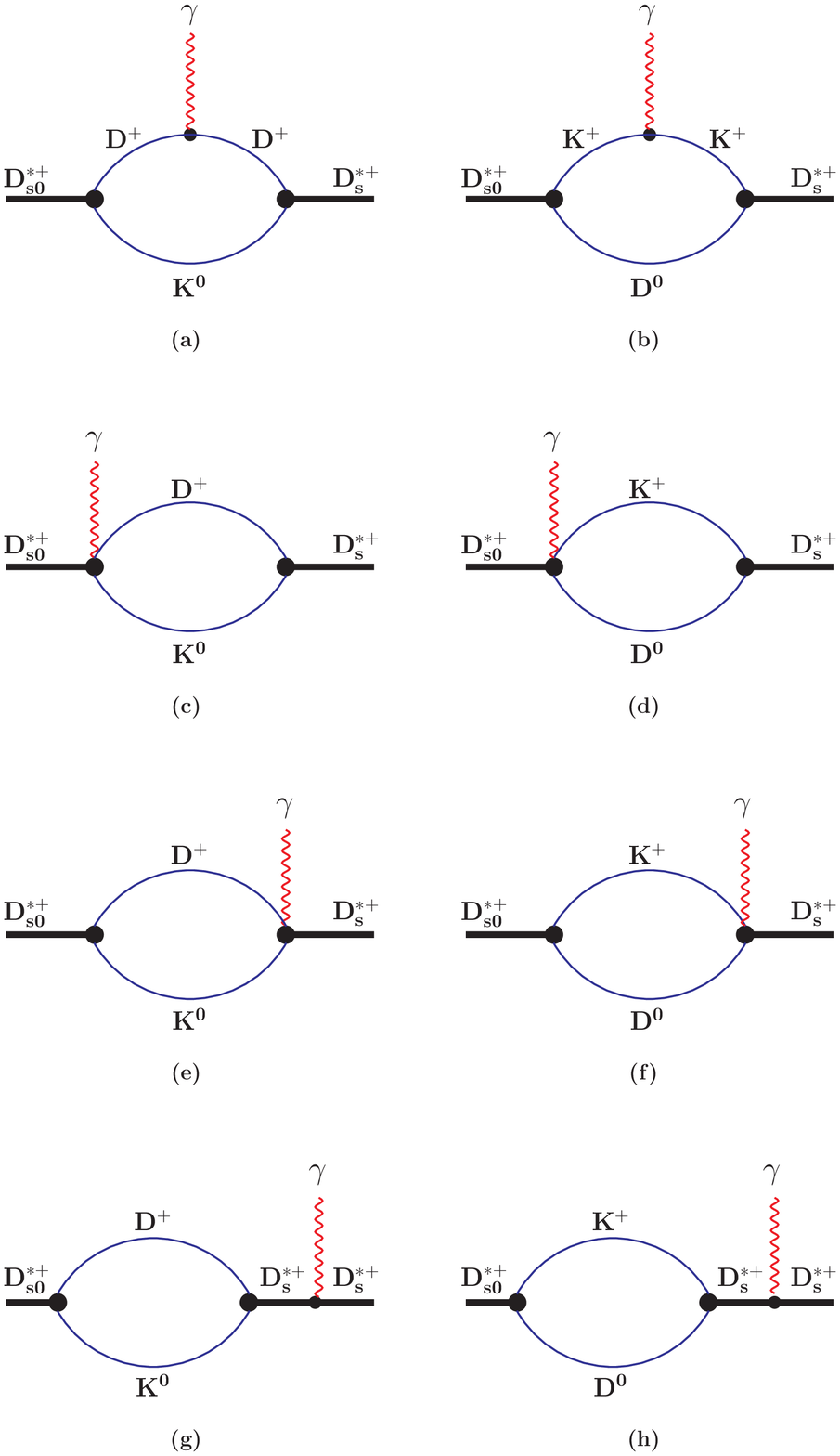}
\figcaption{\label{fig4} Radiative transition $D_{s0}^{\ast +} \to D_{s}^{\ast +} \gamma$}
\end{center}
The other strong coupling vertices are fixed by HHCHPT
or by data, for full details see Ref.~\citep{Faessler:2007gv}.
Similar graphs apply for the strong and radiative transitions of
the $D_{s1} (2460)$ interpreted in the present framework as
a $D^\ast K$ hadronic molecule~\cite{Faessler:2007us}.
Results for the strong decays are
\begin{eqnarray}
\label{eqstrongD}
\Gamma(D_{s0}^{\ast} \to D_s \pi^0) & =& 47 - 112~{\rm keV}\, , \nonumber\\
\Gamma(D_{s1} \to D_s^\ast \pi^0) &=&50 - 79 ~{\rm keV} \, ,
\end{eqnarray}
while for the radiative modes we have
\begin{eqnarray}
\label{eqemD}
\Gamma(D_{s0}^{\ast} \to D_s^\ast \gamma) & =& 0.47 - 0.63~{\rm keV} \, , \nonumber\\
\Gamma(D_{s1} \to D_s \gamma) &=& 2.7 - 3.7 ~{\rm keV} \, .
\end{eqnarray}
The variation in results reflects the uncertainty in the vertex function.
The corresponding ratios of rates with
\begin{eqnarray}
\hspace*{-0.2cm} 
R_{D^\ast_{s0}}&=&\Gamma (D^\ast_{s0}\to D_s^\ast \gamma)
/\Gamma (D^\ast_{s0}\to D_s \pi) =0.01 \nonumber\\
\hspace*{-0.2cm} 
R_{D_{s1}}&=&\Gamma (D_{s1}\to D_s \gamma)
/\Gamma (D_{s1}\to D^\ast _s \pi) = 0.05 
\end{eqnarray}
satisfy qualitatively the present experimental results of
$R_{D^\ast_{s0}}\leq 0.059 $ and
$R_{D_{s1}}=0.44 \pm 0.09 $~\cite{Amsler:2008zz}.
Hence from the present observation of the radiative and strong
decays of the $D^\ast_{s0}$ and $D_{s1}$ an interpretation
as $DK$ and $D^\ast K$ hadron molecules seems feasible.

In Ref.~\citep{Faessler:2007cu} we further investigated the weak
decays $B\to D^{(\ast)} + D^\ast_{s0} (D_{s1})$, where full consistency
with present data is achieved. This further strengthens the presented molecular
picture. We also give predictions~\cite{Branz:2008cb}
for the weak decays involving $f_0(980)$,
which serves as a further indicator for this structure interpretation.

\section{Decay analysis of $X(3872)$}

The $X(3872)$ with quantum numbers $J^{PC} = 1^{++}$ is considered
as a composite state containing both molecular hadronic
and a $c\bar{c}$ component.
We recently showed~\cite{Lee:2009hy} that slight binding of the
$D D^{\ast}$ system can be achieved in a full meson-exchange model.
Following Refs.~\citep{Swanson:2003tb,Dong:2009yp}
we consider the $X(3872)$ state as a superposition
of the dominant molecular $\d\ds$ component, of other hadronic
configurations--$D^{\pm}D^{*\mp}$, $J/\psi\omega$, $J/\psi\rho$,
as well as of the $c\bar c$ charmonium configuration as
\end{multicols}
\ruleup
\begin{eqnarray}
\label{Xstate}
|X(3872)\rangle &=& \cos\theta\Bigg [
\frac{Z_{\d\ds}^{1/2}}{\sqrt{2}}( |\d\dbs \ra + |\ds\db \ra )
+ \frac{Z_{\dppm\dpmps}^{1/2}}{\sqrt{2}}( | \dpp\dpms \ra
+ | \dpm\dpps \ra )  + Z_{\jpsi\omega}^{1/2} | \jpsi \omega \ra
+ Z_{\jpsi\rho}^{1/2} | \jpsi \rho \ra \Bigg ]\nonumber \\
&+&\sin\theta \, c\bar{c} \; .
\end{eqnarray}
\ruledown \vspace{0.5cm}
\begin{multicols}{2}
Here $\theta$ is the mixing angle between the hadronic and the charmonium
components: $\cos^2\theta $ and $\sin^2\theta$ represent the probabilities
to find a hadronic and charmonium configuration, respectively,
for the normalization
\eq
Z_{\d\ds} + Z_{\dppm\dpmps} + Z_{\jpsi\omega} +  Z_{\jpsi\rho} = 1 \,.
\en
For comparison with data we employ values for $Z_{\d\ds}$,
$Z_{\dppm\dpmps}$, $Z_{\jpsi\omega}$ and $Z_{\jpsi\rho}$
as derived in a potential model.
For example, for a binding energy $\epsilon_{\d\ds}
m_{D^0} + m_{D^{\ast 0}} - m_X =0.3$ MeV the estimate of
Ref.~\citep{Swanson:2003tb} provides  
\eq\label{ZH1H2_factors}
Z_{\d\ds}&=& 0.92\,, \hspace*{.25cm}
Z_{\dppm\dpmps} = 0.033\,, \nonumber \\
Z_{\jpsi\omega} &=& 0.041\,, \hspace*{.25cm}
Z_{\jpsi\rho} = 0.006\,.
\en
The coupling of the $X$ to its constituents, once the amplitudes of
Eq. (\ref{ZH1H2_factors}) are chosen, are determined again by
the compositeness condition~\cite{Dong:2009uf,Dong:2009yp,Dong:2008gb}.

A new measurement by the {\it BABAR}
Collaboration gives clear evidence for a strong radiative decay mode
involving the $\psi(2S)$~\cite{Aubert:2008rn}. They indicate the measured
ratio of
\begin{eqnarray}\label{ratio}
\frac{{\cal B}(X(3872)\rightarrow \psi(2S)\gamma)}
{{\cal B}(X(3872)\rightarrow J/\psi\gamma)}=3.5\pm 1.4.
\end{eqnarray}
As known from previous calculations in the molecular approach
the radiative decay width of the $X(3872) \to \psi(2S)\gamma$
is always smaller than the one involving $J/\psi\gamma$.
It is therefore expected that the ratio of Eq.~(\ref{ratio})
gives some constraint on a possible
charmonium component in the $X(3872)$.

In Fig.4 we display the diagrams relevant for the radiative decays
$X\to J/\psi\gamma$ and $X\to\psi(2S)\gamma$.
The last diagram of Fig.4 which
is generated by the $J/\psi V$ $(V=\rho,\omega)$ component of
the $X(3872)$ only contributes to the $J/\psi\gamma$ final state.
\begin{center}
\includegraphics[width=6cm]{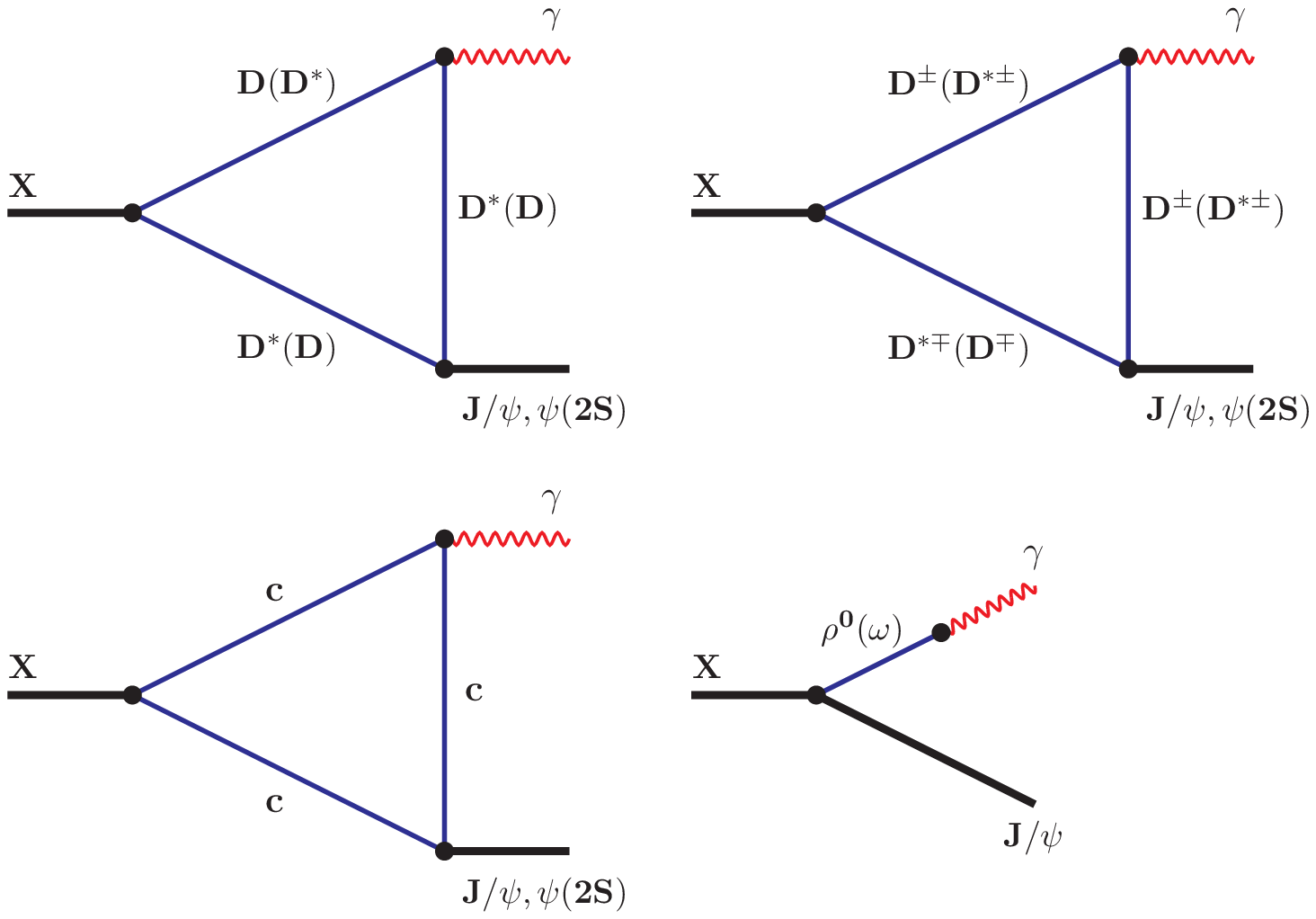}
\figcaption{\label{Fig4} Diagrams contributing to the radiative transitions
$X(3872) \to J/\psi + \gamma$ and $X(3872) \to \psi(2S) + \gamma$.}
\end{center}
Results~\cite{Dong:2009uf} for the radiative decay modes involving the hadronic components
of the $X(3872)$ are contained in Table 1. The value of the binding
energy is set to $0.3$ MeV with the probabilities taken from
Eq.~(\ref{ZH1H2_factors}).

\begin{center}
\tabcaption{ \label{tab1} Radiative decays of hadronic components.}
\footnotesize
\begin{tabular}{|l|l|}
\toprule Hadronic config.
& 
$\Gamma(X(3872) \to \gamma \jp , ~ \gamma \psi(2S))$ keV \\
\hline
 $DD^\ast$ 
& 60 - 120($\jp$) \hfill  0.3 ($\psi(2S)$)  \\
 $J/\psi V$
& 6($\jp$) \hfill  0 ($\psi(2S)$) \\
$DD^\ast$ and $J/\psi V$ 
& 30 - 65 ($\jp$) \hfill  0.3 ($\psi(2S)$)  \\
\bottomrule
\end{tabular}
\end{center}

The inclusion of
the $J/\psi V$ components in the structure of the $X(3872)$
leads to a decrease in the  $X \to J/\psi\gamma$ rate because
$\psi(2s)V$ components are absent in the $X(3872)$.  
Note that taking into account the hadronic components only
the ratio of Eq.~(\ref{ratio}) cannot be explained.
If no $c\bar{c}$ component is present in the $X(3872)$ the decay width for
$J/\psi\gamma$ is about $60$~keV and is much larger than the one for
the $\psi(2S)\gamma$ channel (about $0.3$~keV).
Inclusion of a charmonium component in the X(3872) configuration
leads to the results of Fig.~5 indicating the radiative decay
widths in dependence on the $c\bar c$ admixture.

\begin{center}
\vspace*{.35cm}
\includegraphics[width=5.7cm]{FIG5.eps}
\figcaption{\label{fig5}
Radiative decay widths of the $X(3872)$.
Solid and dotted curves are the results for
the transitions $X \to J/\psi\gamma$ and  $X \to \psi(2S)\gamma$,
respectively.}
\end{center}

A small admixture of a $c\bar{c}$ component
is essential to reproduce the large measured value for
$R_{2S}$~\cite{Dong:2009uf}.
This feature is due to the destructive interference between the
$J/\psi\gamma$ decay amplitudes arising from the
$c\bar{c}$ and the hadronic components.
For mixing values of about $\sin\theta\approx -0.16$ in the case of
$\epsilon =0.3$ MeV the decay width for $\psi(2S)\gamma$ may exceed
the one of $J/\psi\gamma$ and a consistent explanation for the observed
ratio $R_{2S}$ is obtained.

To get estimates~\cite{Dong:2009yp} for the decay widths of $X(3872) \to \jp + h$ with
$h = \pi^+ \pi^- \pi^0, \pi^+ \pi^0, \pi^0 \gamma, \gamma$
we use the results of Ref.~\citep{Braaten:2005ai}, which are based on
the assumption
that these decays proceed through the processes $X$ to $\jp\omega$
and $\jp\rho$.
Here we estimate~\cite{Dong:2009yp} both short and long-distances effects only for the
$X \to \gamma \jp$ decays using our previous result, while
the $X \to \jp + h$ decays
only take into account short--distance effects.
Results for the $\jp$ decay modes are given in Table~2.

\begin{center}
\tabcaption{ \label{tab2} Properties of $X \to \jp + h$ decays.
Numbers in brackets are valid
for explicit values for $Z_{\jpsi\rho}$ and $Z_{\jpsi\omega}$
of Eq.~(\ref{ZH1H2_factors}). Data for ratios
are from Refs.~\citep{Abe:2005ix,Aubert:2008rn}.}
\footnotesize
\begin{tabular}{|l|l|}
\toprule
Quantity &
Nonlocal case \\
\hline
$\Gamma(X \to \jp \pi^+ \pi^-)$, keV

& $9.0 \times 10^3  \, Z_{\jpsi\rho}$ (54.0) \\
\hline
$\Gamma(X \to \jp \pi^+ \pi^- \pi^0)$, keV

& $1.38 \times 10^3  \, Z_{\jpsi\omega}$ (56.6) \\
\hline
$\Gamma(X \to \jp \pi^0 \gamma)$, keV

& $0.23 \times 10^3  \, Z_{\jpsi\omega}$ (9.4) \\
\hline
$\frac{\Gamma(X \to \jp \pi^+ \pi^- \pi^0)}{\Gamma(X \to \jp \pi^+ \pi^-)}$
&  1.05 \\
$= 1.0 \pm 0.4 \pm 0.3$ &\\
\hline
$\frac{\Gamma(X \to \jp\gamma)}{\Gamma(X \to \jp\pi^+\pi^-)}$
 & 0.10 \\
\hskip 0.2cm =$0.14\pm 0.05$; $0.33\pm 0.12$ &\\
\bottomrule
\end{tabular}
\end{center}

Present data for the ratio of rates
$\frac{\Gamma(X \to \jp \pi^+ \pi^- \pi^0)} {\Gamma(X \to \jp \pi^+ \pi^-)}$
indicate a strong isospin violating $\jp \pi^+ \pi^- \pi^0$ decay mode.
Similarly, the measured ratio
$\frac{\Gamma(X \to \jp\gamma)}
{\Gamma(X \to \jp\pi^+\pi^-)}$ points to a large radiative decay channel.
Both ratios of rates can be fully reproduced in the molecular
picture.
Note that all the decay channels considered so far are fed by
the subleading $\jp\omega$, $\jp\rho$ and $c\bar c$ components.
To work out the influence of the leading molecular components
we consider the hadronic
$X \to \chi_{cJ} + (\pi^0, 2\pi)$ and
$X \to \jp + (2\pi, 3\pi)$ decays~\cite{Dong:2009yp}.
Here
the values of $J=0,1,2$ correspond to the $J^P=0^+, 1^+, 2^+$
quantum numbers of the charmonium states. 
Because of the dominance of the $\d\ds$ component
in the transitions of $X$ into charmonium states $\chi_{cJ}$ and pions
we estimate these decays using only this component.
The two-body decays are described
by the $(\d \ds)$ loop diagram shown in Fig.6.

\begin{center}
\includegraphics[width=7cm]{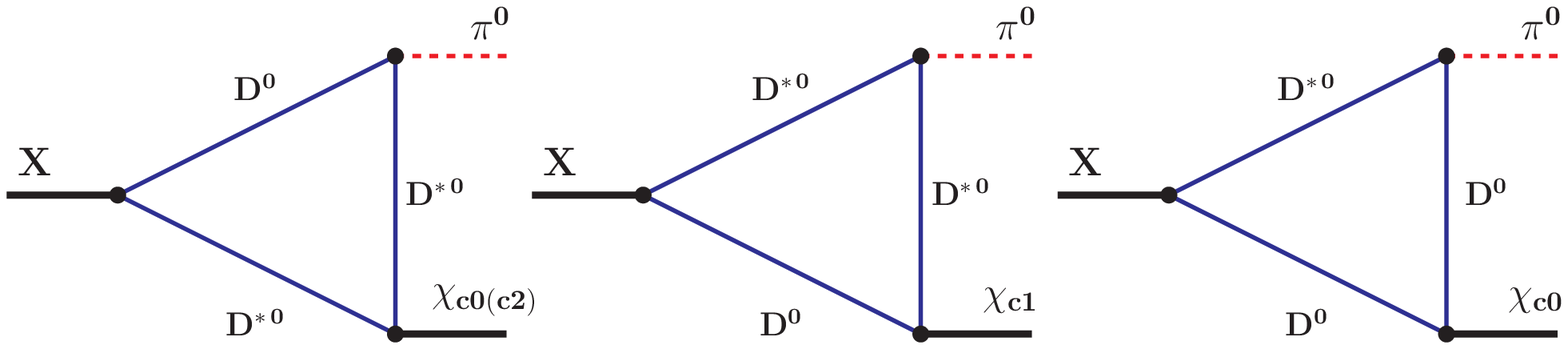}
\figcaption{\label{fig6}
Diagrams contributing to the hadronic transitions
$X(3872) \to \chi_{cJ} + \pi^0$.}
\end{center}

Inclusion of
the charged $(\dppm\dpmps)$ loops gives a further
correction to the decay widths.
The
possible couplings of $D(D^\ast)$ mesons to pions and charmonia states
is taken from HHCHPT.
Similar graphs apply for the three-body
decay modes. Results for the hadronic decays involving
the neutral $(\d \ds)$ component only and the full results
are indicated in Table~3.

\begin{center}
\tabcaption{Properties of $X \to \chi_{cJ} + n\pi$ decays.
Here $Z_D \equiv Z_{\d\ds}$.}
\footnotesize
\begin{tabular}{|l|l|l|}
\toprule
Quantity &  $\d\ds$ loop & $\d\ds +$\\
&&$+ \dpm\dpps$ \\
\hline
            $\Gamma(X \to \chi_{c0} + \pi^0)$, keV & 41.1 $Z_{D}$ (37.8)
                                                   & 61.0 \\
\hline $\Gamma(X \to \chi_{c0} + 2\pi^0)$, eV & 63.3 $Z_{D}$ (58.2) & 94.0 \\
\hline $\Gamma(X \to \chi_{c1} + \pi^0)$, keV & 11.1 $Z_{D}$ (10.2) & 16.4 \\
\hline $\Gamma(X \to \chi_{c1} + 2\pi^0)$, eV & 743 $Z_{D}$ (683.6) & 1095.2 \\
\hline $\Gamma(X \to \chi_{c2} + \pi^0)$, keV & 15 $Z_{D}$ (13.8) & 22.1 \\
\hline $\Gamma(X \to \chi_{c2} + 2\pi^0)$, eV & 20.6 $Z_{D}$ (19.0) & 30.4
 \\
\bottomrule
\end{tabular}
\end{center}

Note that explicit numbers
refer to a binding
energy of $0.3$ MeV with the probabilities taken from
Eq. (\ref{ZH1H2_factors}).
The decay pattern of Table 3 involving pions and $\chi_{cJ}$ states
is sensitive to the leading molecular $(\d \ds)$ component.
These predictions~\cite{Dong:2009yp}
can serve to possibly identify the full hadronic composition of the
X(3872) in
running and planned experiments.

\section{$J/\psi V(=\omega , \phi) $ decays of $Y(3940)$ and $Y(4140)$}

It was suggested in Ref.~\citep{Liu:2009ei}
that both the $Y(3940)$ and $Y(4140)$ are hadronic molecules.
These hadron bound states can have quantum numbers $J^{\rm PC} = 0^{++}$ or
$2^{++}$ whose constituents are the vector charm $D^\ast (D^\ast_s)$ mesons:
\eq\label{M_str}
|Y(3940)\ra &=& \frac{1}{\sqrt{2}} \big(| D^{\ast +} D^{\ast -} \ra+
|D^{\ast 0} \overline{D^{\ast 0}} \ra \big)\,, \nonumber\\
|Y(4140)\ra &=& | D^{\ast +}_s D^{\ast -}_s \ra \,.
\en
For the observed $Y(3940)$ and $Y(4140)$ states
we adopt the convention that the spin and parity
quantum numbers of both states are $J^{\rm PC} = 0^{++}$.
The coupling of the scalar molecular states to their constituents
is again set up by the compositeness condition.
To determine the strong $Y \to J/\psi V$ and two-photon
$Y \to \gamma\gamma$ decays we have to include the couplings
of $D^\ast(D^{\ast}_s)$ mesons to vector mesons
($J/\psi$, $\omega$, $\phi$) and to photons.
The couplings of $J/\psi$, $\omega$, $\phi$ to vector
$D^\ast(D^\ast_s)$ mesons are taken from the HHChPT
Lagrangian~\cite{Wise:1992hn,Colangelo:2003sa}.
The leading-order process
relevant for the strong decays $Y(3940) \to J/\psi \omega$ and
$Y(4140) \to J/\psi \phi$ is the diagram of Fig.7
involving the vector mesons $D^\ast$ or $D^\ast_s$ in the loop.

\begin{center}
\includegraphics[width=4.5cm]{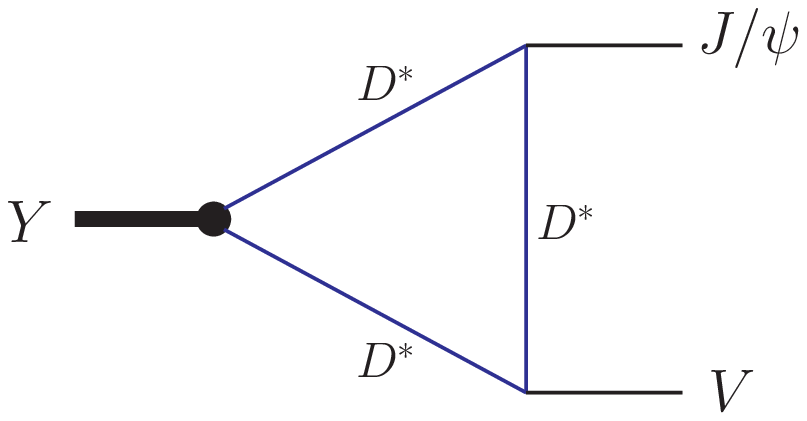}
\figcaption{\label{fig7}
Leading order diagram for $Y \to J/\psi V$ decay.}
\end{center}

We also consider the radiative $Y(3940)/Y(4140) \to \gamma\gamma$
decay widths~\cite{Branz:2009yt},
which serve as a prediction for possibly identifying the molecular structure
of these Y states.
\begin{center}                                                                  
\includegraphics[width=6cm]{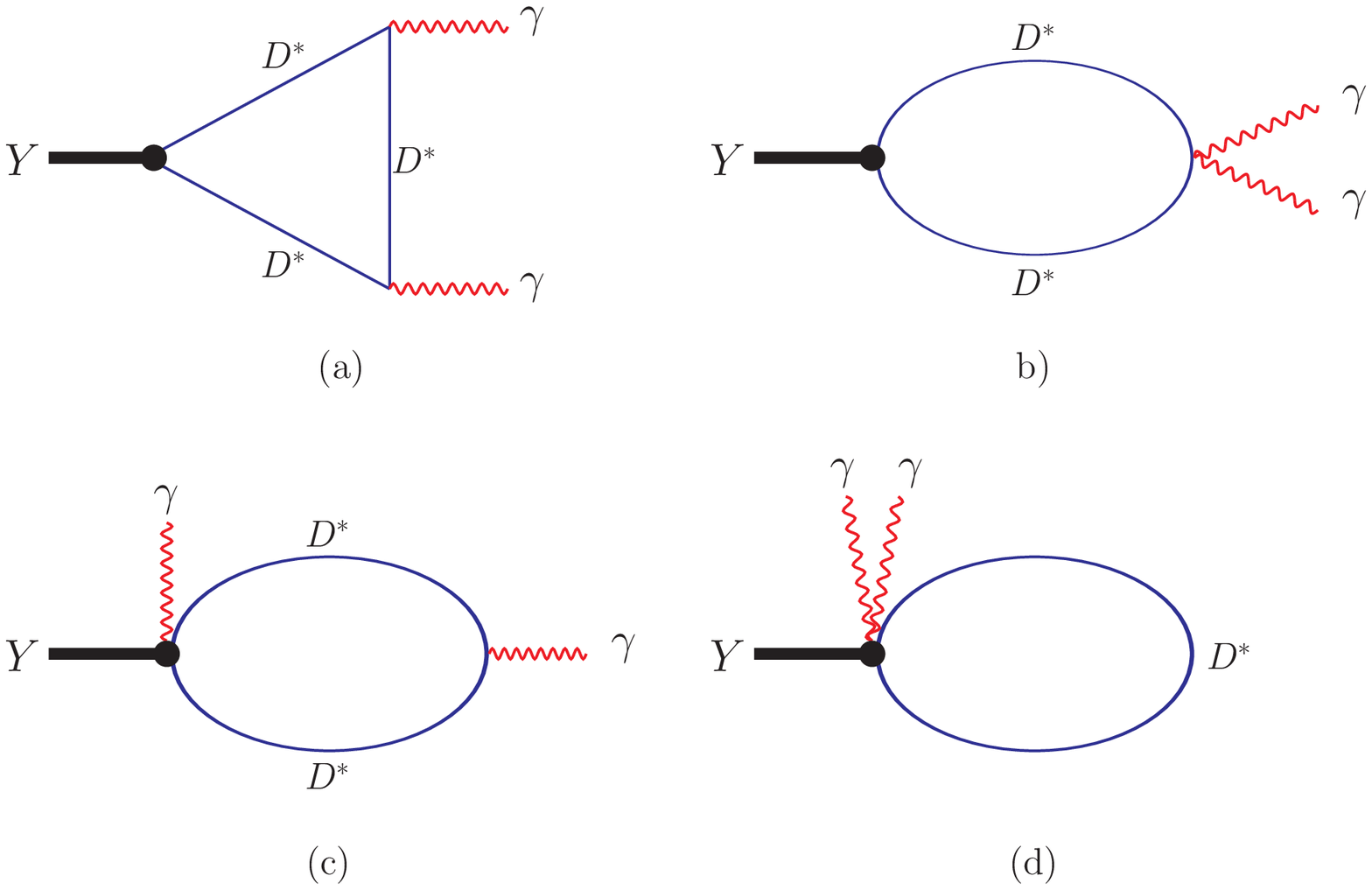}  
\figcaption{\label{fig8}                                                        
Diagrams relevant for the radiative decay $Y \to \gamma \gamma$.}               
\end{center}
The relevant diagrams which arise from coupling of the charged $D^{\ast \pm}(D^{\ast \pm}_s)$
mesons to photons and from gauging the strong interaction Lagrangian are displayed in Fig.8.
The numerical results for the quantities characterizing the strong
$J/\psi V$ ($V = \omega, \phi$) and radiative $2\gamma $  decays
of $Y(3940)$ and $Y(4140)$ are contained in Table 4.

\begin{center}
\tabcaption{Decay properties of $Y(3940)$ and $Y(4140)$.}
\footnotesize
\begin{tabular}{|l|c|c|}
\toprule
Quantity                            & $Y(3940)$ & $Y(4140)$ \\
\hline
$\Gamma(Y \to J/\psi V)$, MeV       & 5.47 $\pm$ 0.34 & 3.26 $\pm$ 0.21 \\
\hline
$\Gamma(Y \to \gamma\gamma)$, keV   & 0.33 $\pm$ 0.01 & 0.63 $\pm$ 0.01 \\
\hline
$\displaystyle{R = \frac{\Gamma(Y \to \gamma\gamma)}
{\Gamma(Y \to J/\psi V)}} \times 10^4$
& $0.61 \pm 0.06$ & $1.93 \pm 0.16$ \\
\bottomrule
\end{tabular}
\end{center}

The predictions of $\Gamma( Y(3940) \to J/\psi \omega )=5.47$~MeV
and $\Gamma (Y(4140) \to J/\psi \phi )=3.26$~MeV for the observed
decay modes are sizable and fully consistent with the
upper limits set by present data on the total widths.
The result for $\Gamma( Y(3940) \to J/\psi \omega )$ is also
consistent with the lower limit of about 1 MeV~\cite{Eichten:2007qx}.
Values of a few MeV for these decay widths naturally arise in the
hadronic molecule interpretation of the $Y(3940)$ and $Y(4140)$,
whereas in a conventional charmonium interpretation the $J/\psi V$
decays are strongly suppressed by the Okubo, Zweig and Iizuka
rule~\cite{Eichten:2007qx}.
Further tests of the presented scenario concern the two-photon decay
widths, which we predict to be of the order of 1 keV~\cite{Branz:2009yt}.
Results for the strong $J/\psi $ decays are quite
similar for the $J^{\rm PC} = 2^{++}$ assignment, which cannot be ruled out
at this point.

\section{Conclusions}

The approach to hadron molecules based on the compositeness condition
constitutes a consistent field theoretical tool.
The determination
of the decay properties of the open charm mesons 
$D_{s0}^*(2317)$,
$D_{s1}(2460)$
and the hidden charm mesons X(3872), Y(3940) and Y(4140) in comparison
to present data is fully in line
with a molecular interpretation of these states. We also give further
predictions of decay properties which can be tested in future experiments.
Current results are very encouraging in ultimately identifying
hadronic molecules in the meson spectrum.

\acknowledgments{We thank Yubing Dong, Yong-Liang Ma and Sergey Kovalenko
for extensive collaborations on the topic of hadron molecules.  
V.E.L. is on leave from the Department of Physics,
Tomsk State University, 634050 Tomsk, Russia.}

\end{multicols}

\vspace{-2mm}
\centerline{\rule{80mm}{0.1pt}}
\vspace{2mm}

\begin{multicols}{2}

\end{multicols}

\vspace{5mm}

\clearpage

\end{document}